\documentstyle[twocolumn,aps,amsfonts,epsfig]{revtex}

    \def\z{\noindent}  
 \def\Box{{\hfill\hbox{\enspace${\sqre}$}} \smallskip}
    \def\sqr#1#2{{\vcenter{\vbox{\hrule height .#2pt
                             \hbox{\vrule width .#2pt height#1pt \kern#1pt
                                   \vrule width .#2pt}
                             \hrule height .#2pt}}}}
 \def\sqre{\mathchoice\sqr54\sqr54\sqr{4.1}3\sqr{3.5}3}

\begin{document}
\wideabs{ \title{Decay Versus Survival of a Localized
State Subjected to Harmonic Forcing: Exact Results}

\author{A.Rokhlenko, O. Costin, J. L. Lebowitz\footnote{Also
Department of Physics}} \address{Department of Mathematics,
Rutgers University\\ Piscataway, NJ 08854-8019}

\maketitle
\begin{abstract}
                                    
We investigate the survival probability of a localized 1-d quantum 
particle subjected to a time dependent potential of the form 
$rU(x)\sin{\omega t}$ with $U(x)=2\delta (x-a)$ or   $U(x)=
2\delta(x-a)-2\delta (x+a)$. The particle is initially in a bound
state produced by the binding potential $-2\delta (x)$. 
We prove that this probability goes to zero as $t\to\infty$ for 
almost all values of $r$, $\omega$, and $a$.
The decay is initially exponential followed by a $t^{-3}$ law  
if $\omega$ is not close to resonances and $r$ is small;
otherwise the exponential disappears and Fermi's golden rule fails. 
For exceptional sets of parameters $r,\omega$ and $a$ the survival 
probability never decays to zero, corresponding to the Floquet
operator having a bound state. We show similar behavior even in the 
absence of a binding potential: permitting a free particle to be 
trapped by harmonically oscillating delta function potential.

\medskip

\z PACS: 03.65.Db; 03.65.Ge; 32.80.Fb

\end{abstract}}

\narrowtext

{\bf 1. Introduction}
\vskip0.3cm
Quantum systems subjected to strong external
time dependent fields often show very complex behavior, e.g. the
ionization probability of an atom may be a complicated  
function of the frequency, amplitude, pulse shape and other parameters of the
field [1-4]. Such phenomena, which go beyond conventional
perturbation theory in the field strength ($r$ here), are readily seen
in numerical solutions of the time dependent Schr{\"o}dinger
equation. There are also various approximate analytic methods which
reproduce many experimental results [5,6,7], but there is however no
rigorous theory of such phenomena even for model systems. In this
paper we describe new exact results for a toy model which has 
both bound and continuum eigenstates, subjected to a
harmonically oscillating potential. They reveal a very rich
structure for the time evolution of even very simple quantum system.
In particular the transition from a bound
state to the continuum is seen to be much more complex than the simple
exponential decay obtained from conventional perturbation theory
via Fermi's golden rule [2].

To obtain exact results we need to consider simplified
model systems. In particular we cannot treat (for the present time) 
realistic description of the interaction between radiation and
matter, such as the dipole approximation. We note however, that a 
comparison of our earlier results on an even simpler version of this 
model with experiments on the strong field ionization of
Rydberg atoms [8] showed some surprising similarity between the two.
We interpret this as an indication of a certain universality in the 
Schr{\"o}dinger evolution of a system with bound and continuum 
spectrum subjected to time dependent external forces and that our 
model retains some of its behavior.
\vskip0.6cm
{\bf 2. The model}

The important idealizations in our model are: 1) Space is 
one-dimensional, 2) The
"internal" potential creating the bound state is given by an
attractive delta function at the origin, 3) The interaction with the external
field has the form $\eta (t)U(x)$ where $\eta$ is periodic in time
with a rectangular envelope, and
4) $U(x)$ is given by one or more delta functions at different
locations on the $x$-axis. The first two assumptions are quite common
for modeling short range binding potentials [9,10] and should not
affect greatly the basic physics of the ionization process.
Assumption 3 means that we do not consider situations [11] where there is
some "ramping" in turning the external field on and off. This should
not be too serious when the pulse is of a long duration compared to
the period of the field which is the case we are concerned with here. 
Assumption 4 on the other hand clearly makes the interaction in our
model very different from the real interactions between radiation and 
matter: dissociation due to electromagnetic fields are described 
approximately by a dipole interaction of the form $U(x)=x$. 
Unfortunately we have not been able to obtain exact results for this 
case beyond those described in [11]. The only feature of the dipole 
interaction we are able to mimic is the spatial symmetry. 

Using suitable units in which $\hbar =2m=1$ ($m$ is the particle mass)
the time evolution of our system is
given by the Schr{\"o}dinger equation $$
i{\partial\over \partial t}\psi(x,t)=\left [-{\partial^2\over \partial x^2}-
2\delta(x)+U(x)\eta(t)\right ]\psi(x,t),\eqno(1)$$
where $\eta(t)=r\sin{\omega t}$ and the parameters $r,\omega$
represent the amplitude and frequency of the time dependent potential.
The spatial structure of the external potential will be 
taken in two forms: $U_1(x)=2\delta (x-a)$ and $U_2(x)=2[\delta (x+a)-
\delta (x-a)]$; $U_2$ has the symmetry of the dipole
interaction. The factor $2$ in front of the binding potential is
chosen so that the unique bound state of the unperturbed system 
is $u_b(x)=e^{-|x|}$ with binding energy $E_0=\omega_0=1$. 
Eq.(1) is to be solved subject to the initial condition 
$\psi(x,0)=u_b(x)$. We can readily extend our methods to more
general sums of delta functions. Our main interest is
in the survival probability of the bound state at time
$t$: $|\theta(t)|^2=|\langle \psi(x,t)u_b(x)\rangle |^2$.

The case $U_1(x)$ with $a=0$, which corresponds the parametric 
perturbation of the binding potential, was treated in [12,13]. We showed 
there that $|\theta(t)|^2$
has both exponential and power law parts which are well
separated only when the strength of perturbation $r$ is small. This
was true for all $\omega$ away from resonances 
($\omega\not\approx N^{-1},\ N$ an integer), with $\omega <1$ corresponding
to ionization via ``multi-photon'' processes. We also obtained there 
non-monotonic dependence of the escape rate on $r$ and
$\omega$. Somewhat to our surprise we found qualitative (and even
semi-quantitative) agreement between the
predictions of this model about resonance behavior of the survival 
probability of localization and some experimental observations on the
ionization of Rydberg atoms by strong microwave fields [8]. It was also
proved in [13] that when $\eta(t)$ is a sum of a finite number of
harmonics, ($\sum_{j=1}^M A_je^{ij\omega t}+ $ complex conjugate), then
the survival probability $|\theta(t)|^2\to 0$ as $t\to\infty$ for any
$M<\infty ,\ A_M\neq 0$. There are however very special infinite 
sequences $A_j$, given explicitly in [13], for which we proved that 
the system never ionizes fully $|\theta(t)|^2\not\to 0$. 

Here we show that the situation is quite different and much richer
when we consider $U_1(x)$ with $a\neq 0$ or $U_2(x)$.
(In physical units the position of perturbation corresponds
to $\hbar a/\sqrt{2mE_0}$).
In particular we prove that for $\eta(t)=r\sin\omega t$ there exist
two-dimensional manifolds in the space of the three parameters $\omega,
r, a$ on which $|\theta(t)|^2\not\to 0$ as $t\to \infty$. (We shall take 
without loss of generality $\omega, r, a$ positive.) This means that
while $|\theta(t)|^2\to 0$ for almost all parameter values of the 
forcing, ``exceptional'' cases can also be constructed quite readily. 
This does not occur for fixed $\omega$ and $a$ if $r$
is small enough and is thus outside conventional perturbation theory. 
We find in addition that when $\omega$ is very close to a 
resonance, $\omega\approx N^{-1}+$ dynamic Stark shift, 
then the decay may not have the exponential part predicted by the golden
rule no matter how small $r$ is even when $|\theta(t)|^2\to 0$.

We can also consider the case when
there is no binding potential at all, i.e. the term 
$-2\delta(x)$ is absent in (1). In this case we have for
$\eta(t)=0$ a free particle, which when initially localized
in the vicinity of the origin will diffuse away: the probability
of being in any fixed region decaying as $t^{-1}$. On the other hand 
the perturbations with special values
of $a,\omega ,r$ can make the particle stay localized for all time.
\vskip0.4cm
{\bf 3. Results for $U_2(x)$}

We give here an outline of the proof which follows along the lines
presented in detail for the case $a=0$ in [13].
Expanding $\psi (x,t)$ in terms of the eigenfunctions of the
unperturbed Hamiltonian $H_0=-{d^2\over d x^2}-
2\delta(x)$, we write $$
\psi (x,t)=\theta (t)e^{-|x|+it}+\int_{-\infty}^\infty \Theta (k,t)
u(k,x)e^{-ik^2t}dk,\eqno(2)$$
where the initial conditions are $\theta(0)=1,\ \Theta (k,0)=0$ and 
the explicit expression [12] for the continuum states are $$
u(k,x)={1\over \sqrt{2\pi}}\left (e^{ikx}-{e^{i|kx|}\over i+i|k|}
\right ),\ \ -\infty<k<\infty .$$
Substituting (2) into
(1) we obtain $\psi (x,t)$ in the form of a functional of $\psi (a,t)$
and $\psi (-a,t)$ while $\theta (t)$ can be written as$$
\theta(t)=1+2ie^{-a}\int_0^t\eta (s)[Y^+(s)-Y^-(s)]ds,\eqno(3)$$
where $Y^{\pm}(t)=e^{-it}\psi(\pm a,t)$. We now express $\Theta (k,t)$
in terms of $Y^{\pm}(t)$ too, take 
$x=\pm a$ in (2) and obtain a coupled pair of integral equations
$$Y^\pm(t)\!=\!e^{-a}\!+\!\int_0^t\!\eta (s)\!\left[K^\pm
(\!t-\!s)Y^+(s)\!-\!K^\mp (\!t\!-s)Y^-(s)\right ]ds.\eqno(4)$$
The Laplace transforms of the kernels $K^\pm (t)$ are $$
k^-(p)={e^{-2a\sqrt{1-ip}}\over 
\sqrt{1-ip}-1},\ k^+(p)={1+k^-(p)\over \sqrt{1-ip}},$$ 
with the choice of branch $\sqrt{1-ip}\to 1$ when $p\to 0$.
For $\eta(t)=r\sin \omega t$, letting $y^\pm(p)$ be the Laplace 
transforms of $Y^\pm(t)$ and setting $f(p)=y^+(p)-y^-(p)$, eq.(3)
yields
$$\theta(t)={re^{-a}\over 2\pi i}\int_C
{e^{pt}\over p}[f(p-i\omega)-f(p+i\omega)]dp\eqno(5)$$
with integration along a contour $C$ which goes from
$-i\infty$ to $i\infty$ in the right half plane avoiding $p=0$ by
a small semi-circle in the left half plane.

{\it Survival of bound state:} The survival
probability $|\theta(t)|^2$ is determined (see (5)) by the analytic
structure of $y^\pm (p)$. Setting $y_n^\pm =y(p+i\omega n),\ 
k^\pm_n=k^\pm(p+i\omega n)$ eqs.(4) turn into
the recurrence relations for the vectors $y_n=\{y^+_n,y^-_n\},
\ n\in {\bf Z}$ 
$$y_n={e^{-a}\over p+i\omega n}{1 \atopwithdelims () 1}-{ir\over 2}
\pmatrix{k^+_n&-k^-_n\cr
k^-_n&-k^+_n\cr}
(y_{n-1}-y_{n+1}),\eqno(6)$$
\vskip-0.3cm
\noindent  
where $y_n,\ k^\pm_n$ may be viewed as functions
of the parameter $p$ in the strip $\Im(p)\in [0,\omega )$.  
The poles of $y(p)$ are in the left half-plane at $\xi_0+i\omega n,
\ \xi_0\leq 0$, and the branch points
at $p=-i-in\omega,\ n\in {\bf Z}$ (the latter ones are inherited from
$k^\pm_n(p)$). After making  horizontal cuts at $p=x+in\omega,\
-\infty<x=\Re {p}\leq 0$ we push the contour $C$ in (5) along the
branch cuts into the left half
plane. The poles then contribute a series of residues with the
common exponential factor $e^{\xi_0 t}$ while the integrals around the
cuts generate a contribution in the form of a series in terms of
$t^{-j-1/2},\ j\geq 1$ (see [12]). The imaginary part of $\xi_0$, 
$\Im\xi_0$, is identified as the dynamical Stark shift 
[2,7] of the resonance frequency and 
$\Gamma=-2\Re\xi_0$ is the decay exponent in the initial stage of 
evolution when $r$ is small and the exponential and polynomial parts
of $|\theta(t)|^2$ may be separated.

If $\Re\xi_0=0$, i.e. the poles lie on the imaginary
axis, then $\theta(t)\not\to 0$ as $t\to\infty$. This can
happen when on the imaginary $p$ axis the homogeneous recurrence, $$
z_n=-{ir\over 2}
\pmatrix{k^+_n&-k^-_n\cr
k^-_n&-k^+_n\cr}
(z_{n-1}-z_{n+1}),\eqno(7)$$
\vskip-0.3cm
\noindent 
associated with (6), has non-trivial solutions which decay
sufficiently rapidly as $n\to\pm\infty$, i.e. $\sum |z_n|^2<\infty$.
This is a manifestation of the Fredholm alternative [14]. We show now that
unlike the case $a=0$ treated in [12,13] such solutions
though non-generic are possible for $U_2(x)$ and
for $U_1(x)$ with $a\geq 1/2$. 

Setting $p=ig$ (with a real $g$) we
construct a particular solution of (7) for $\omega >1$.
It is clear that if $z_j=z_{j+1}=0$ then all successive $z_n$ will be
zero too until the matrix in (7) becomes degenerate. We set 
$z_n=0$ for all $n\leq 0$ and 
require the determinant $(k^+_0)^2-(k^-_0)^2$ to vanish, which
allows $z_1\neq 0$, in particular $z^+_1=z^-_1$. This implies
\vskip -0.5cm
$$a\sqrt{-1-g_0}=\pi N,\ N=1,2,...,\eqno(8)$$
\vskip -0.1cm\noindent
where the parameter $g_0\in (-\omega ,-1)$ represents the ``binding energy''
of a localized state produced by the perturbation (Floquet state [6,7]).
For $n\geq 1$ all $k^\pm_n$ are real positive and the
matrices in (7) are non-degenerate. By inverting
them it is easy to show that $z^-_n=(-1)^{n+1} z^+_n$ and obtain a scalar
recurrence for $z^+_n,\ n\geq 0$. Using a new variable $\rho_n=
iz^+_{n+1}/z^+_n$  the recurrence (7) takes the form
\vskip -0.5cm
$$\rho_n={2\over r k_n}-{1\over \rho_{n-1}},\ \rho_1={2\over rk_1}, 
\eqno(9)$$

\vskip -0.2cm
\noindent
where $n>1$ and $k_n=k^+_n+(-1)^{n+1}k^-_n$.

A careful analysis for the case $N=1$ in (8) shows that the decaying 
solutions of (9) can be constructed with a unique $r=r(a,\omega)$. Two 
inequalities,  
$$k_1k_2\geq 2k_3(2k_3-k_2),\ \ k_1k_2\geq k_3(4k_3-k_2),\eqno(10)$$
which are necessary and sufficient for the existence of
solutions of (9), specify regions in the $a,\omega$, plane. Relations (10)
can always be satisfied if we choose $g_0+\omega \ll 1$ which makes
$k_1$ large. The second of them gives the upper bound on
$\omega$ which becomes very strict, $0<\omega +g_0\ll 1$ when 
$-g_0\sim 1$, i.e. $a$ is large, see (8).
The interval where the stabilizing $r$ is located can be specified
too and we can prove that for an arbitrary frequency  
$\omega >1$ there is an interval of $a=\pi /\sqrt{-1-g_0}$ with $g_0$
in $(-\omega ,-1)$ where, for a particular choice
of amplitude $r=r(a,\omega)$ the system does not ionize
completely. Instead
$\theta (t)\to e^{ig_0t}F(\omega_s t)$, where $\omega_s$ is the
stabilizing frequency of the perturbation and
$F$ is a periodic function with period $2\pi$.
\vskip0.6cm
{\bf 4. Results for $U_1(x)$}

This type of stabilization of the bound state takes place for the
perturbation with the potential $U_1(x)$ too, but only when $a\geq
1/2$. In this case there
will again be a 2-d manifold in $a,r,\omega$ variables for which
the bound state is stabilized and $\theta(t)\to e^{ig_0t}F(\omega_s t)$,
a quasi periodic function of $t$. We also computed $\theta (t)$ numerically
for this model by solving the integral equation for $Y(t)$
and the most representative curves are shown in Fig.1. 
The slowest decay 
on the time interval $0<t\leq 160\pi/\omega_0\approx 500$ was obtained
for $\omega =1.12$ which is close to the value of $\omega_s\approx
1.089...$, evaluated by constructing the decaying 
solution of (9), and pushing it to as large $n$ as we can within the 
accepted precision. $|\theta(t)|^2$  
near the Stark shifted resonance, $\omega=1.2$, has no
interval in which the decay is exponential in contrast with such a
decay observed for $\omega =1.25$ and $\omega =0.8$.
\vskip -2.3cm
\begin{figure}
\hskip 2cm \epsfig{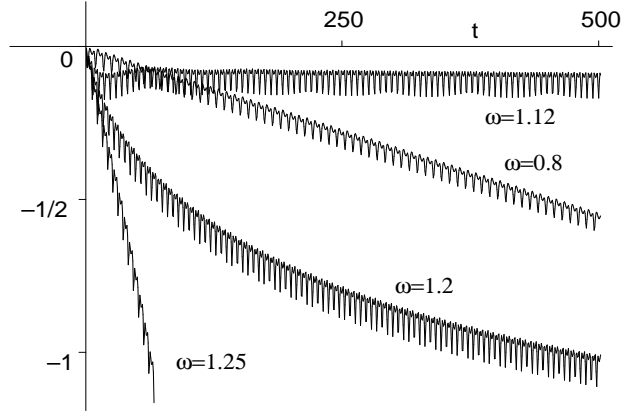}
\hskip0.5cm
\caption{Plot of $\log_{10}|\theta(t)|^2$ for $a\approx 0.59,\ r=1$.}

\end{figure}

The ripples on the curves in Fig.1 have the frequency 
of the perturbation $\omega$ and the modulation is due to
beats among $\omega$, $\omega_0$ and $g_0$. 

A perhaps simplistic explanation of the stabilization is maybe 
to view it as some kind of bouncing of the trapped 
particle between the delta potential wells. Note that the 
eigenfrequency $g_0$ of
the new bound state is a function (see (8)) of $a$ only, but the
amplitude $r$ or the frequency $\omega$ must be  
fine tuned as functions of $g_0$ to prevent the leaking out of 
the particle wave function.
\vskip0.6cm
{\bf 5. Time decay of the bound state}

For the model with $U_1(x)$ the recurrence 
(6) has a scalar form and its solution can be
written in terms of continued fractions which converge quite rapidly
when $r$ is small. For $\omega>1$, neglecting terms of order of $r^4$ and
higher, one may truncate the recurrence around each
$n$ by taking $y_{n+m}=0$ if $|m|\geq 2$. The solution of the 
truncated system for $n=0$, which gives the main contribution to
$\theta(t)$ has a pole at $\xi_0=O(r^2)$, that solves the
equation
\vskip -0.5cm
$$1+{r^2\over 4}k^+_0(\xi_0)\big [k^+_1(\xi_0)+k^+_{-1}(\xi_0)\big ]=O(r^4).
\eqno(11)$$
\vskip-0.2cm
\noindent
Using (11) the contour integration in (5), where $Y^-\equiv 0$, yields
\vskip-0.7cm 
$$\theta(t)\approx e^{\xi_0(\omega)t}+{\omega
e^{i(\omega -1-\Delta)t+i\pi /4}\Re \xi_0(\omega)\over \sqrt{\pi}
[(\omega^2 -(1+\Delta)^2][(\omega -1-\Delta)t+1]^{3/2}},\eqno(12)$$
where $\Re \xi_0(\omega)=-r^2\lambda(\omega)\sqrt{\omega
-1-\Delta},\ \Im\xi_0(\omega)=\Delta =r^2\sigma(\omega)$, 
and $\lambda(\omega),\sigma(\omega)\not =0$ 
are of order $e^{-2a}$ when $a$ is large. $\Delta$ represents the Stark
shift [6,7].

The survival probability $|\theta(t)|^2$ has initially an exponential regime
where it decays 
as $e^{-\Gamma t}$, with $\Gamma =-2\Re \xi_0$ proportional to $r^2$ in
agreement with Fermi's golden rule.
As $t$ increases an increasingly important role is
played by transitions back to the bound state with probability 
proportional to the density near the origin as given by
the second term of (12). As a result, for $t\gg \Gamma^{-1}$, the 
survival probability follows the power law decay $|\theta(t)|^2\sim
t^{-3}$, (see also [4,13,15,16]). The mathematical origin of this
power law is the square root branch point at the bottom of the
continuous spectrum.
Note that this is much faster than when an initially localized 
free particle is permitted to evolve. The probability of it
remaining localized then decays as $t^{-1}$ [17]. 
\vskip -2.0cm
\begin{picture}(00,00)(00,00)
\put(84,-75){ $\scriptstyle 10^6\ \ \ \ \ \ \ \ \ \ \ \ \ \ \ \ \ 2\cdot 10^6$}
\end{picture}
\begin{figure}
\hskip -0cm\epsfig{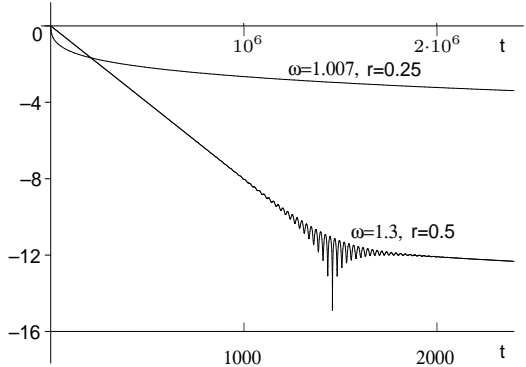}
\vskip 0.4cm
\caption{$\log_{10}|\theta(t)|^2$ for two regimes of decay (plot of
Eqs.(12),(14), note the different time scales and drastically different 
rates of decay). The upper curve for resonant $\omega$ does not follow
Fermi's rule. The lower curve with the regular
behavior (exponential with a $t^{-3}$ tail) is farther from resonance.}
\end{figure}

{\bf 6. Resonances}

Near the one photon resonance, $\omega \gtrsim 1$ (11) implies  
$$\xi_0(\omega)+r^2\lambda (1)\sqrt{\omega -1-\xi_0(\omega)}=
ir^2\sigma(1)+O(r^4).\eqno(13)$$
The solution of (13) gives $\Delta =O(r^2)$ and $\Gamma =o(r^2)$
whose dependence on $r$ is determined by the order in $r$ of $\omega
-1$, but in the case when $\omega$ is very close to $1+\Delta$ we 
cannot separate anymore the
contributions in (3) because the poles of $y(p)$ are too close to the
branch points. Integration in (5) yields$$
\theta(t)\approx {e^{i\epsilon t+i\pi/4}\over \pi}\int_{-\infty}^\infty
{e^{-tr^4\lambda^2(1)x^2/4}x^2dx\over [x^2-h(a)]^2+ix^2},\eqno(14)$$
\vskip-0.2cm
\noindent
where $h(a)=O(1)$ when $a$ is not large.
For  $r^4t\gg 1$ eq. (14) implies
$|\theta(t)|^2\to t^{-3}$, but the initial exponential regime does not
exist even for small $r$, see curves in Fig.2, which are constructed
using (12)-(14). (Compare with [4,15,16]).

{\it ``Multiphoton'' resonances}.
Let us compute the decay exponent when $\omega <1$ which
corresponds to a ``multi-photon ionization'' [6,8,12] for our simple
model. To
locate the singular point $p$ (which should be somewhere in the left
half-plane very close to the imaginary axis) we require the homogeneous
recurrence (a scalar analogue of (7)) to be solvable by a sequence
$z_n$ such that $|z_n|\to 0$ when $|n|\to\infty$. Using (9) and the
continued fractions we represent $\rho_0$ in terms of 
$\rho_n^{-1}$ for $n<0$ and  $\rho_n$ when $n>0$. Both representations are
rapidly convergent for $r\ll 1$ and the condition of their matching
yields explicitly the solution for $\xi_0$ in the lowest order $\Delta =
O(r^2)$ as before and $\Gamma =O( r^{2N})$. Zeros of $k_{-n}$ 
change the order of the
decay exponent $\Gamma$ of the multiphoton ionization from $r^{2N}$ to
$r^{2N+2}$ which drastically slows down the the ionization rate. 
The stabilization can also be expected for $\omega\approx 1/2,1/3,...$
and therefore when $\omega <1/2$ we practically cannot see the
decay if $t$ is not extremely large. 
\vskip0.6cm
{\bf 7. Trapping of free particle} 

Let us remove in (1) the binding
potential and take the wave function$$
\psi(x,0)=\int_{-\infty}^\infty F(k)u(k,x)dk,\ \ \int_{-\infty}^\infty 
|F(k)|^2dk=1,$$
which describes a localized particle in the vicinity of the origin. Here
$u(k,x)=(2\pi )^{-1/2}e^{ikx}$ are the free eigenfunctions. For $\eta(t)=0$
the initial state evolves as$$
\psi^0(x,t)=\int_{-\infty}^\infty F(k)u(k,x)e^{-ik^2t}dk\to t^{-1/2},
\eqno(15)$$
when $t\to\infty$. Using in (1) the expansion in
terms of $u(k,x)$ and Laplace transform ($\psi(x,t)\to \tilde 
\psi(x,p)$) we obtain the infinite set of equations$$
\tilde\psi_n(x)=\tilde\psi^0_n(x)+$$
\vskip-1cm
$$\eqno(16)$$
\vskip-1cm
$${r\over 2}[ k^+_n(x)(y^+_{n-1}-y^+_{n+1})-
k^-_n(x)(y^-_{n-1}-y^-_{n+1})],$$
where $F_n(x)$ means $F(x,p+i\omega n)$, $y^\pm_n=\tilde \psi (\pm
a,p+\omega n)$, and
\vskip-0.5cm
$$k^\pm(x,p)=i{e^{i|x\mp a|\sqrt{ip}}\over \sqrt{ip}},\
\sqrt{ip}\to i\ {\rm as}\ p\to i.\eqno(17)$$
\vskip-0.2cm
Setting $x=a$ and $x=-a$ in (16) we obtain a recurrence similar to (6)
with vectors $\{\psi_n^0(a),\psi_n^0(-a)\}$.
If on the imaginary axis the homogeneous recurrence (7) 
has a properly decaying solution for some $a,\omega ,r$ then all 
$y_n$ have poles at $p=ig(a)+i\omega n$ respectively and 
therefore their inverse Laplace transforms which represent 
$\psi(\pm a,t)$ do not vanish as $t\to\infty$. For a given $\omega$
the requirement that the determinant $(k^+_0)^2-(k^+_0)^2$ vanishes
gives the parameter $g_0\in (-\omega ,0)$ as a
function of $a$: $g_0=-(\pi N/a)^2$ ($N$ is an integer and clearly
$a>\pi/\sqrt{\omega}$). Setting $z_n=0$ for all $n\leq 0$ and
$z^+_1=z^-_1$, we can again invert the matrices in (7), obtain
(9) and repeat the previous computation. The
explicit form of the coefficients,
\vskip-0.5cm
$$k_n={1+(-1)^{n+1}e^{-2a\sqrt{g_0+\omega n}}\over \sqrt{g_0+\omega
n}},\eqno(18)$$
\vskip-0.2cm\noindent
implies here too a possibility to find a decaying sequence $\rho_n$ and
a rapidly decreasing set of $z^\pm_n\sim r^n\omega^{-n/2}/\sqrt{n!}$, as 
$n\to\infty$.

The poles of $y_n$ develop by (17) into poles of $\tilde \psi (x,p)$
at the same points $p=ig_0+i\omega n$. Therefore as $t\to\infty$ the wave
function $\psi(x,t)$ will survive near the origin an can be
represented as a series related to poles of (16) in Laplace space
$$e^{ig_0t}\sum_{n=1}^\infty \left [e^{-|x-a|\sqrt{g_0+\omega n}}
Q_n(\omega t)+e^{-|x+a|\sqrt{g_0+\omega n}}P_n(\omega t)\right ],$$
where the coefficients $P_n,Q_n$ are periodic functions which decay
rapidly as $n$ tends to $\infty$.
\vskip0.6cm
{\bf 8. Concluding remarks}

Our results show the richness of the structure exhibited by a simple toy
model driven externally in the presence of a continuum. The survival
probability can change greatly, including trapping in a localized
state, as the parameters of the external forcing are varied.
While not all the features of this simple model can be expected to 
be mirrored by real atoms driven by electromagnetic fields we
believe that some features are rather universal. In particular the power
law decay [16] and the Fermi golden rule violation at resonances
are expected to occur quite generally. Even if the location of
resonance is not on the real axis of the energy plane but has a small 
imaginary component and therefore the localized state is slowly
decaying, the power law tail can compete with the exponent on the
whole observable time interval. 
\vskip0.6cm
{\bf Acknoledgments}

We thank R. Barker, R. Schrader and A. Soffer for useful
comments. Research supported by AFOSR Grant \# F49620-01-0154 and
NSF Grants \# 0103807, 0100495.


\vskip-0.6cm
\begin{thebibliography}{15}

\bibitem{[1]} {\it Atoms in Intense Laser Fields}, Ed. M. Gavrila,
{\it Adv. Atom. Mol. Opt. Phys. Supplement 1}, (Academic Press, San Diego
1992); {\it Adv. Atom. Mol. Opt. Phys.}, Eds. B. Bederson and H. Walther,
(Academic Press, San Diego 1995); {\it Super-Intense Laser-Atom
Physics}, Eds. B. Piraux, A. L'Huillier,
K. Rza{\.z}evsky, (Plenum Press, New York 1993).

\bibitem{[2]} C.Cohen-Tannoudji, J.Dupont-Roc, and G.Grynberg, {\it
Atom-Photon Interactions} (Wiley, New York, 1992). 

\bibitem{[3]} {\it Multiphoton Ionization of Atoms}, Eds. S. L. Chin and
P. Lambropoulos, (Academic Press, Toronto, New York 1984);

\bibitem{[4]} S.R.Wilkinson, C.F.Bharucha, M.C.Fischer,
K.W.Madison, P.R.Morrow, O.Nlu, B.Sundaram, and M.G.Raizen, 
Nature {\bf 387}, 575 (1997).

\bibitem{[5]} R.M.Potvliege and R.Shakeshaft, Phys.Rev.A
{\bf 40}, 3061 (1989); S.Chelkowsky, A.D.Bandrauk, and P.B.Corkum,
PRL {\bf 19}, 2355 (1990).

\bibitem{[6]} J.N.Bardsley and M.J.Comella, Phys.Rev.A {\bf 39},
2252 (1989); M.S.Pindzola and M.D{\"o}rr, Phys.Rev.A {\bf
43}, 439 (1991).

\bibitem{[7]} N.B.Delone and V.P.Krainov, {\it Multiphoton
Processes in Atoms}, Springer-Verlag, Berlin-New York (1994);
A.Maquet, Shih-I Chu and W.P.Reinhardt, Phys. Rev. A {\bf
27}, 2946 (1983); W.R.Salzman, Phys.Rev.A {\bf 10}, 461 (1974);
L.Pan, K.T.Taylor, and C.W.Clark, Phys.Rev.A {\bf 43}, 6272 (1991);
G.Scharf, K.Sonnenmoser, and W.F.Wreszinski, 
Phys.Rev.A  {\bf 44}, 3250 (1991); A.Buchleitner, D.Delande,
J.Zakrzewski, R.N.Mantegna, M.Arndt, and H.Walther, PRL {\bf
75}, 3818 (1995).

\bibitem{[8]} P.M.Koch, Acta Physica Polonica A {\bf 93} No. 1, 105
(1998); P.M.Koch and K.A.H.van Leeuwen, Phys.Reports {\bf
255}, 289 (1995); T.J.Bensky, G.Haeffler, and R.R.Jones, PRL {\bf
79}, 2018 (1997); F.Benvenuto, G.Casati, and D.L.Shepelyansky, 
Phys.Rev.A {\bf 45}, R7670 (1992); F.Benvenuto, G.Casati, and 
D.L.Shepelyansky, Phys.Rev.A {\bf 47}, R786 (1993).

\bibitem{[9]} S.Geltman, J.Phys. B {\bf 10}, 831 (1974); 
S.M.Susskind, S.C.Cowley, and E.J.Valeo, Phys.Rev. A {\bf 42}, 3090
(1994); Yu.N.Demkov and V.N.Ostrovskii, {\it Zero Range Potentials
and Their Application in Atomic Physics} (Plenum, 1988).   

\bibitem{[10]} S. Albeverio, F. Gesztesy, R. Hoegh-Krohn, and 
H. Holden, {\it Solvable models in quantum mechanics},
Springer-Verlag, New York-Berlin (1988).

\bibitem{[11]} $\mbox{A.Fring, V.Kostrykin, and R.Schrader, J. Phys. B: At.}$ 
Mol. Opt.Phys. {\bf 29}, 5651 (1996);
A.Fring, V.Kostrykin, and R.Schrader,
J.Phys.A: Math.Gen. {\bf 30}, 8559 (1997).

\bibitem{[12]} O.Costin, J.L.Lebowitz, and A.Rokhlenko, J.Phys.A
{\bf 33}, 6311 (2000).

\bibitem{[13]} O.Costin, R. D. Costin, J.L.Lebowitz, and A.Rokhlenko,
C.M.P. {\bf 221}, 1 (2001).

\bibitem{[14]} M.Reed and B.Simon, {\it Methods of Modern
Mathematical Physics} (Academic Press, New York, 1972).

\bibitem{[15]} G.Garcia-Calderon, J.L.Mateos, and M.Moshinsky,
PRL {\bf 74}, 337 (1995); A.Buchleitner, D.Delande and J.-C.Gay,  
J.Opt.Soc.B {\bf 12}, 505 (1995). A.Soffer and M.I.Weinstein,
Jour.Stat.Phys. {\bf 93}, 359 (1998). 

\bibitem{[16]} A.Soffer and M.I.Weinstein, J.Stat.Phys. {\bf 93}, 
359 (1998). 

\bibitem{[17]} A.Rokhlenko and J.L.Lebowitz, J.Math.Phys.
{\bf 41}, 3511 (2000).


\end{thebibliography}
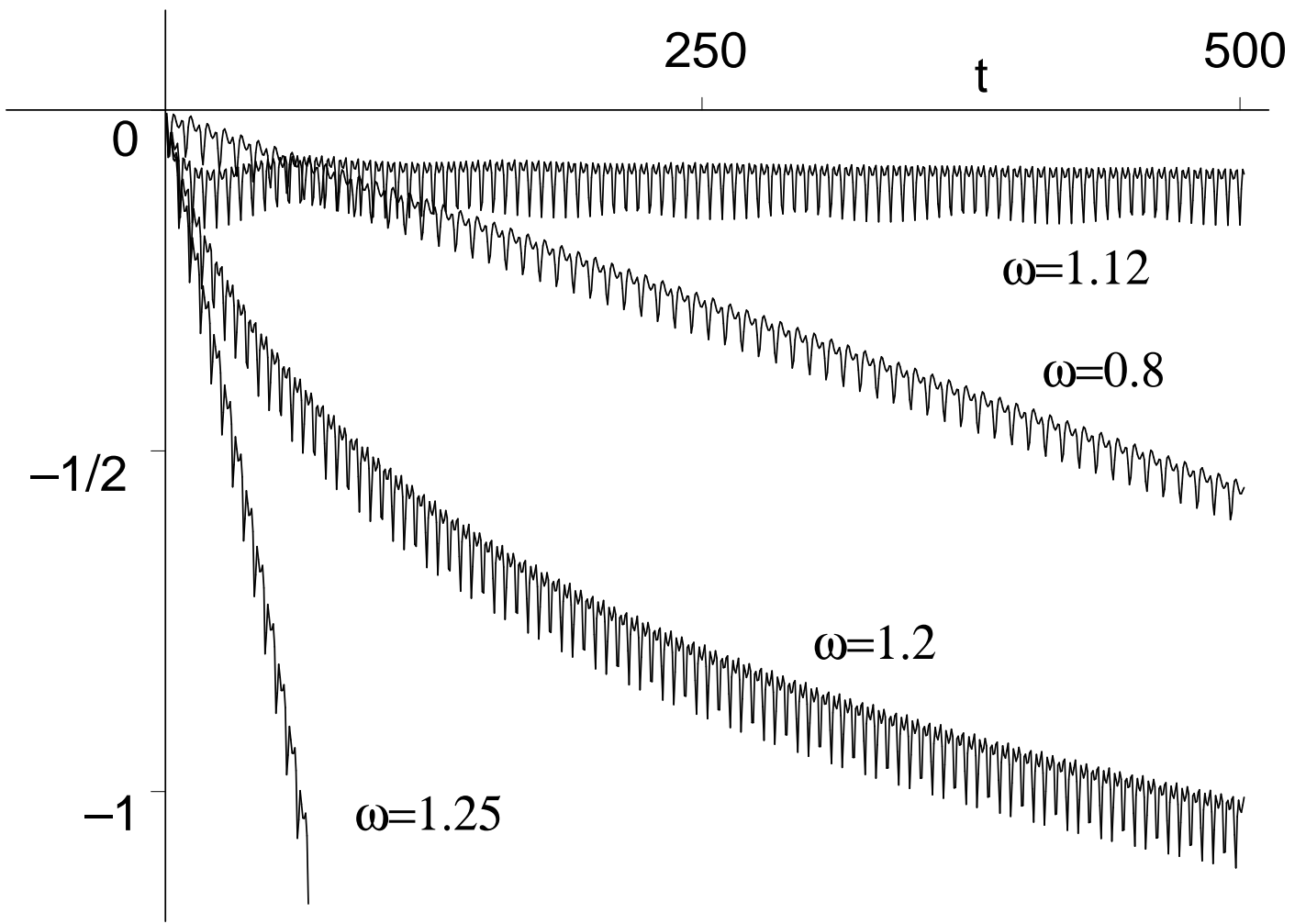
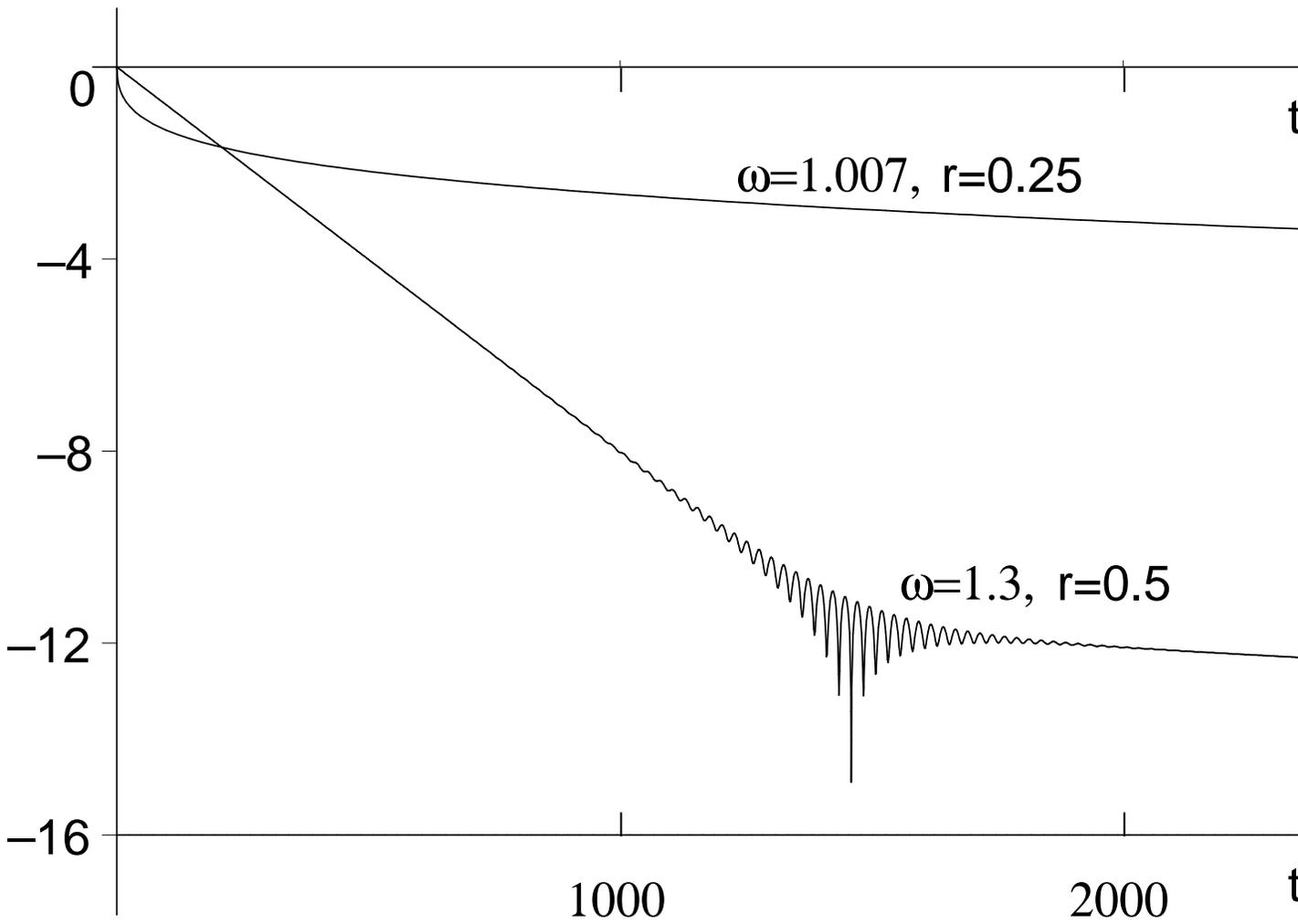
\end{document}